\documentclass[slac_one]{revtex4}
\usepackage{graphicx}
\usepackage{fancyhdr}
\begin{document}

\title{Recent Results on $\psi(3770)$ Physics at BES-II}

\author{H. L. Ma (For BES Collaboration)}
\affiliation{IHEP, CAS, Beijing 100049, P.R.China}

\begin{abstract}
About 33, 6.5 and 1.0 pb$^{-1}$ of $e^+e^-$ annihilation data were,
respectively, taken around the center-of-mass energies of $\sqrt s=$
3.773 GeV, at $\sqrt s=$ 3.650 GeV and at $\sqrt s=$ 3.6648 GeV with the BES-II
detector at the BEPC collider. By analyzing these data sets, we measure the
branching fraction for $\psi(3770)\to$ non-$D\bar D$; observe an anomalous line
shape of the cross section for $e^+e^-\to$ hadrons in the energy region
from 3.650 to 3.872 GeV; and measure the line shapes of the $D^+D^-$,
$D^0\bar D^0$ and $D\bar D$ production and the ratios of the production
rates of $D^+D^-$ and $D^0\bar D^0$ in $e^+e^-$ annihilation at $\psi(3770)$
resonance. We also search for $\psi(3770)$ decay into exclusive light hadron
processes containing $K^0_S$ or $\pi^0\pi^0$ mesons in the final states.
\end{abstract}

\maketitle

\thispagestyle{fancy}

\section{Introduction}
The $\psi(3770)$ is the lowest mass charmonium resonance above the
open charm pair $D\bar D$ production threshold. Traditional theories
expect that it decays almost entirely into pure $D\bar D$ pairs.
However, before the measurements from BES and CLEO Collaborations,
there is a long standing puzzle that the observed cross section
$\sigma_{\psi(3770)}^{\rm obs}$ for $\psi(3770)$ production is not
saturated by the observed cross section $\sigma_{D\bar D}^{\rm obs}$
for $D\bar D$ production at the
$\psi(3770)$ peak \cite{rzhc}. 
Recently, CLEO
Collaboration measured the $e^+e^-\to\psi(3770)\to$ non-$D\bar D$
cross section to be $(-0.01\pm0.08^{+0.41}_{-0.30})$ nb \cite{prl96_092002}.
But, BES measured the branching
fraction for $\psi(3770)\to$ non-$D\bar D$ by analyzing several different
data samples and different methods to be $(14.7\pm3.2)\%$
\cite{plb641_145,prl97_121801,prd76_122002,plb659_74,pdg2008} with assumption that there is
only one simple $\psi(3770)$ in the energy region between 3.700 and 3.872
GeV. While, up to now, the sum of the measured branching fractions for
$\psi(3770)\to$ exclusive non-$D\bar D$ decays is not more than 2\%
\cite{hepnp28_325,plb605_63,prl96_082004,prl96_182002,prd74_031106,prd74_012005}.
To better understand this situation, we examine the line shape
of the cross section for $e^+e^-\to$ hadrons in the energy region from 3.650
to 3.872 GeV, measure the line shapes of the $D^+D^-$, $D^0\bar D^0$ and $D\bar D$
production and the ratios of the production rates of $D^+D^-$ and $D^0\bar
D^0$ in $e^+e^-$ annihilation at $\psi(3770)$ resonance, and search for
$\psi(3770)\to$ exclusive light hadron final states.

These measurements are made by analyzing about 33, 6.5 and 1.0 pb$^{-1}$
of $e^+e^-$ annihilation data sets taken around the center-of-mass energies
of $\sqrt s=$ 3.773 GeV, at $\sqrt s=$ 3.650 GeV and at $\sqrt s=$ 3.6648 GeV
with the BES-II detector at the BEPC collider. The data sets taken around
$\sqrt s=$ 3.773 GeV include the data sets of 17.3 pb$^{-1}$ taken at
$\sqrt s=$ 3.773 GeV, the precision cross section scan data sets taken
in March 2001, during March to April 2003 and during December 2003 to
January 2004.

\section{Measurements of the branching fraction for $\psi(3770)\to$
non-$D\bar D$ decays \cite{plb641_145,prl97_121801,prd76_122002,plb659_74}}
We measure the branching fraction for $\psi(3770)\to$ non-$D\bar D$
decays by analyzing several different data samples
and different methods with assumption that there is only one simple
$\psi(3770)$ in the energy region between 3.700 and 3.872 GeV.
The measured branching fractions for
$\psi(3770)\to D^0\bar D^0$, $D^+D^-$, $D\bar D$ and non-$D\bar D$
decays are compared in Tab. \ref{tab:brdd}. After the $\psi(3770)$ resonance was
discovered for more than thirty years, Particle Data Group 2008
\cite{pdg2008} gives the branching fractions for $\psi(3770)\to
D^0\bar D^0$, $D^+D^-$ and $D\bar D$ decays for the first time.
They are $B[\psi(3770)\to D^0\bar D^0]=(48.7\pm3.2)\%$,
$B[\psi(3770)\to D^+D^-]=(36.1\pm2.8)\%$ and
$B[\psi(3770)\to D\bar D]=(85.3\pm3.2)\%$,
This indicates that the branching fraction for $\psi(3770)\to$ non-$D\bar D$
decays is $(14.7\pm3.2)\%$.

These measurements imply that $\psi(3770)$ could substantially decay into
non-$D\bar D$ final states, which might greatly challenge the traditional
theories. Otherwise, there may exist some other effects in the energy region
around $\psi(3770)$ resonance which are responsible for the large branching
fraction for $\psi(3770)\to$ non-$D\bar D$ decays. On the other hand,
we also need to try to search for more charmless decays of $\psi(3770)$
\cite{plb650_111,plb656_30,epjc52_805}.

\begin{table*}[htbp]
\begin{center}
\caption{\label{tab:brdd}
The measured branching fractions for $\psi(3770)\to D^0\bar D^0$,
$D^+D^-$, $D\bar D$ and non-$D\bar D$ decays.}
\begin{tabular}{|c|c|c|c|c|} \hline
$B[\psi(3770)\to]$(\%) &$D^0\bar D^0$     &$D^+D^-$          &$D\bar D$        &non-$D\bar D$      \\ \hline
Ref.\cite{plb641_145}  &$49.9\pm1.3\pm3.8$&$35.7\pm1.1\pm3.4$&$85.5\pm1.7\pm5.8$&$14.5\pm1.7\pm5.8$\\
Ref.\cite{prl97_121801}&$46.7\pm4.7\pm2.3$&$36.9\pm3.7\pm2.8$&$83.6\pm7.3\pm4.2$&$16.4\pm7.3\pm4.2$\\
Ref.\cite{prd76_122002}&-&-&-&$13.4\pm5.0\pm3.6$\\
Ref.\cite{plb659_74}   &-&-&-&$15.1\pm5.6\pm1.8$\\ \hline
PDG 2008\cite{pdg2008}  &$48.7\pm3.2$&$36.1\pm2.8$&$85.3\pm3.2$&$14.7\pm3.2$\\
\hline
\end{tabular}
\end{center}
\end{table*}

\section{Anomalous line shape of the cross section for $e^+e^-\to$ hadrons
in the energy region from 3.650 to 3.872 GeV \cite{prl101_102004}}

\begin{figure*}[hbtp]
\centering
\includegraphics[width=8cm,height=6cm]{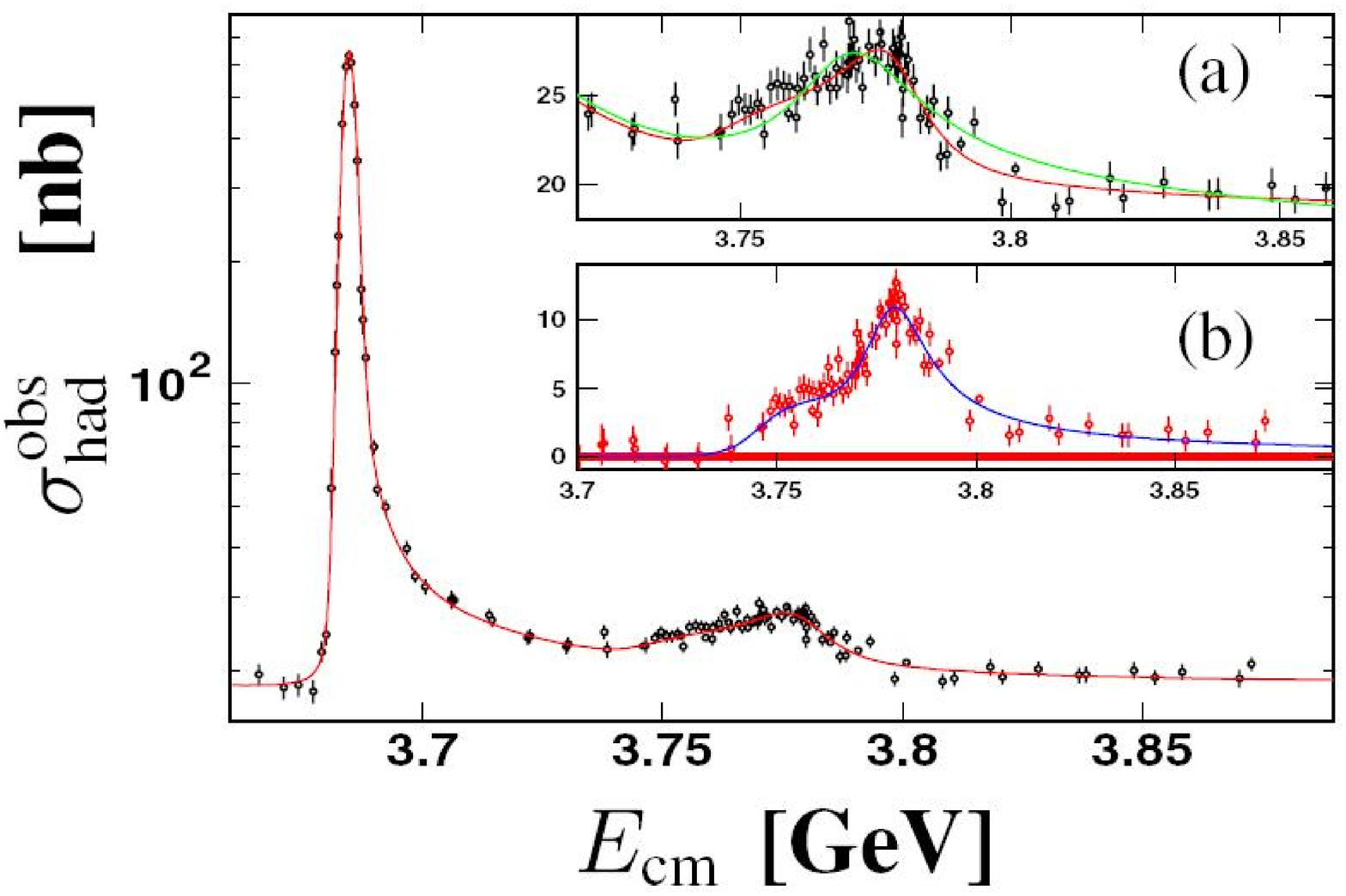}
\caption{The observed cross sections for $e^+e^-\to$ hadrons versus the
nominal center-of-mass energies for solution 2.}
\label{fig:crs_hads}
\end{figure*}

To better understand why the measured branching fraction for $\psi(3770)\to$
non-$D\bar D$ is substantially larger than 2\%, we examine the line shape
of the cross section for $e^+e^-\to$ hadrons in the energy region from 3.650
to 3.872 GeV. By analyzing the precision cross section scan data sets taken
in March 2003 and during December 2003 to January 2004, we obtain the
measured observed cross sections for $e^+e^-\to$ hadrons versus the nominal
center-of-mass energies, as shown in Fig. \ref{fig:crs_hads}. In the figure,
we can see that the slope of the high-energy side of the peak is
substantially larger than that of the low-energy side.
This phenomenon is inconsistent with the traditional expectation under
the assumption that there is only one simple $\psi(3770)$ resonance in
this energy region.

To investigate this situation, we fit the measured observed cross sections for
$e^+e^-\to$ hadrons with the following solutions, respectively.
Firstly, we suppose that there are two amplitudes and ignore the possible
interference between them.
Secondly, we suppose that there are two amplitudes completely interfering with
each other.
Thirdly, we assume that there are two amplitudes of $G(3900)$
\cite{babar,belle} and $\psi(3770)$ resonance interfering with each other.
Finally, we assume that there is only one simple $\psi(3770)$ resonance.
The fitted results are summarized in Tab. \ref{tab:anomalous_crs_had}.
The details about the fits can be found in Ref. \cite{prl101_102004}.
By comparing the fitted results, we can obtain the better hypothesis to
describe the anomalous line shape of the cross sections for $e^+e^-\to$
hadrons in the energy region from 3.700 to 3.872 GeV.
The signal significance for the two structure hypotheses are $7.0\sigma$
and $7.6\sigma$ for solution 1 and solution 2. The significance of the
interference between the two amplitudes is $3.6\sigma$.
Fig. \ref{fig:crs_hads} shows the fit to the observed cross sections
for $e^+e^-\to$ hadrons for solution 2.

\begin{figure*}[hbtp]
\centering
\includegraphics[width=8cm,height=6cm]{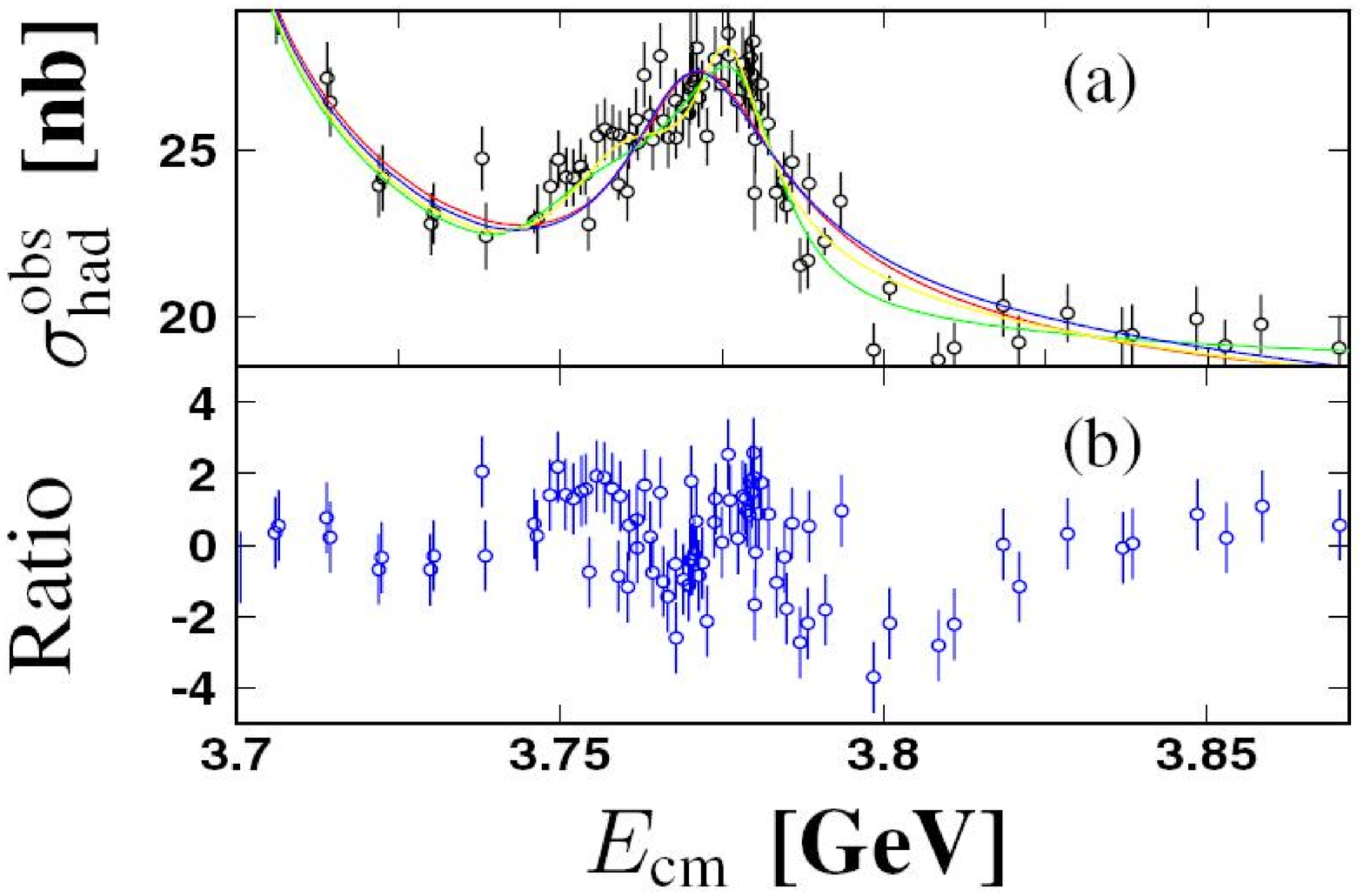}
\caption{
(a) The observed cross sections for $e^+e^-\to$ hadrons versus the
nominal center-of-mass energies;
(b) the ratio of the residual to error of observed cross section.}
\label{fig:com_fit}
\end{figure*}

Fig. \ref{fig:com_fit}(a) shows the fit to the observed cross sections
for $e^+e^-\to$ hadrons for solution 3, compared with the fits for solution
1 and solution 2.
Fig. \ref{fig:com_fit}(b) shows the ratio of the residual between the
observed cross section and the fitted value for the one $\psi(3770)$
amplitude hypothesis to the error of the observed cross section,
which indicating that there is more likely some new structure in addition to
$\psi(3770)$ resonance.

\begin{table*}[htbp]
\begin{center}
\caption{\label{tab:anomalous_crs_had}
The fitted results, where $M$, $\Gamma^{\rm tot}$ and $\Gamma_{ee}$ are the
mass, the total width and the leptonic width of resonance(s),
$\sigma_G$ is the standard deviation of $G(3900)$, $\phi$ is the phase
difference between the two amplitudes and AM stands for amplitude(s). ndof
denotes the number of degrees of freedom. The upper $^{S1}$, $^{S2}$ and
$^{S3}$ represent the Solution 1, Solution 2 and Solution 3, respectively.}
\begin{tabular}{|c|c|c|c|c|} \hline
Quantity & Two AM$^{S1}$ & Two AM$^{S2}$ & One AM & $\psi(3770$ and
$G(3900)$$^{S3}$ \\ \hline
$\chi^2$/(ndof) & 125/103 = 1.21 & 112/102 = 1.10 & 182/106 = 1.72 & 170/104
=1.63 \\
$M_{\psi(3686)}$[MeV]   & $3685.5\pm0.0\pm0.5$ & $3685.5\pm0.0\pm0.5$ &
$3685.5\pm0.0\pm0.5$ & $3685.5\pm0.0\pm0.5$\\
$\Gamma^{\rm tot}_{\psi(3686)}$[keV]   & $312\pm34\pm1$ & $311\pm38\pm1$ &
$304\pm36\pm1$ & $293\pm36\pm1$\\
$\Gamma^{ee}_{\psi(3686)}$[keV]   & $2.24\pm0.04\pm0.11$ &
$2.23\pm0.04\pm0.11$ &
$2.24\pm0.04\pm0.11$ & $2.23\pm0.04\pm0.11$\\
$M_1$[MeV]   & $3765.0\pm2.4\pm0.5$ & $3762.6\pm11.8\pm0.5$ &
$3773.3\pm0.5\pm0.5$ & $3774.4\pm0.5\pm0.5$\\
$\Gamma^{\rm tot}_1$[eV]   & $28.5\pm4.6\pm0.1$ & $49.9\pm32.1\pm0.1$ &
$28.2\pm2.1\pm0.1$ & $28.6\pm2.3\pm0.1$\\
$\Gamma^{ee}_1$[eV]   & $155\pm34\pm8$ & $186\pm201\pm8$ & $260\pm21\pm8$ &
$264\pm23\pm8$\\
$M_2$[MeV]   & $3777.0\pm0.6\pm0.5$ & $3781.0\pm1.3\pm0.5$ &... &
3943.0(fixed)\\
$\Gamma^{\rm tot}_2[eV]$   & $12.3\pm2.4\pm0.1$ & $19.3\pm3.1\pm0.1$ & ... &
...\\
or $\Gamma^G$[MeV]   & ... & ... &  & 54(fixed)\\
$\Gamma^{ee}_2$[eV]   & $93\pm26\pm9$ & $243\pm160\pm9$ & ... & ...\\
or C  & ... & ... &  ... & 0.243(fixed)\\
$\phi$[deg] & ... & $(158\pm334\pm5)$ & ... & $(150\pm23\pm5)$\\
$f$ & $0.4\pm5.6\pm0.6$ & $5.2 \pm 2.5 \pm0.6$ & $0.0\pm0.5\pm0.6$ &
$0.0\pm1.2\pm0.6$\\
\hline
\end{tabular}
\end{center}
\end{table*}

\section{Measurements of the line shapes of $D\bar D$ production and the
ratios of the production rates of $D^+D^-$ and $D^0\bar D^0$ in $e^+e^-$
annihilation at $\psi(3770)$ resonance \cite{plb668_263}}

To investigate what on earth is responsible for the large branching
fraction for $\psi(3770)\to$ non-$D\bar D$ decays which is beyond the expectation
by the traditional theories, we measure the line shapes of the $D^+D^-$,
$D^0\bar D^0$ and $D\bar D$ production and the ratios of the production
rates of $D^+D^-$ and $D^0\bar D^0$ in $e^+e^-$ annihilation at
$\psi(3770)$ resonance. These measurements are also helpful for
the understanding the anomalous line shape of the cross section for
$e^+e^-\to$ hadrons in the energy region from 3.650 to 3.872 GeV.

\begin{figure*}[hbtp]
\centering
\includegraphics[width=8cm,height=6cm]{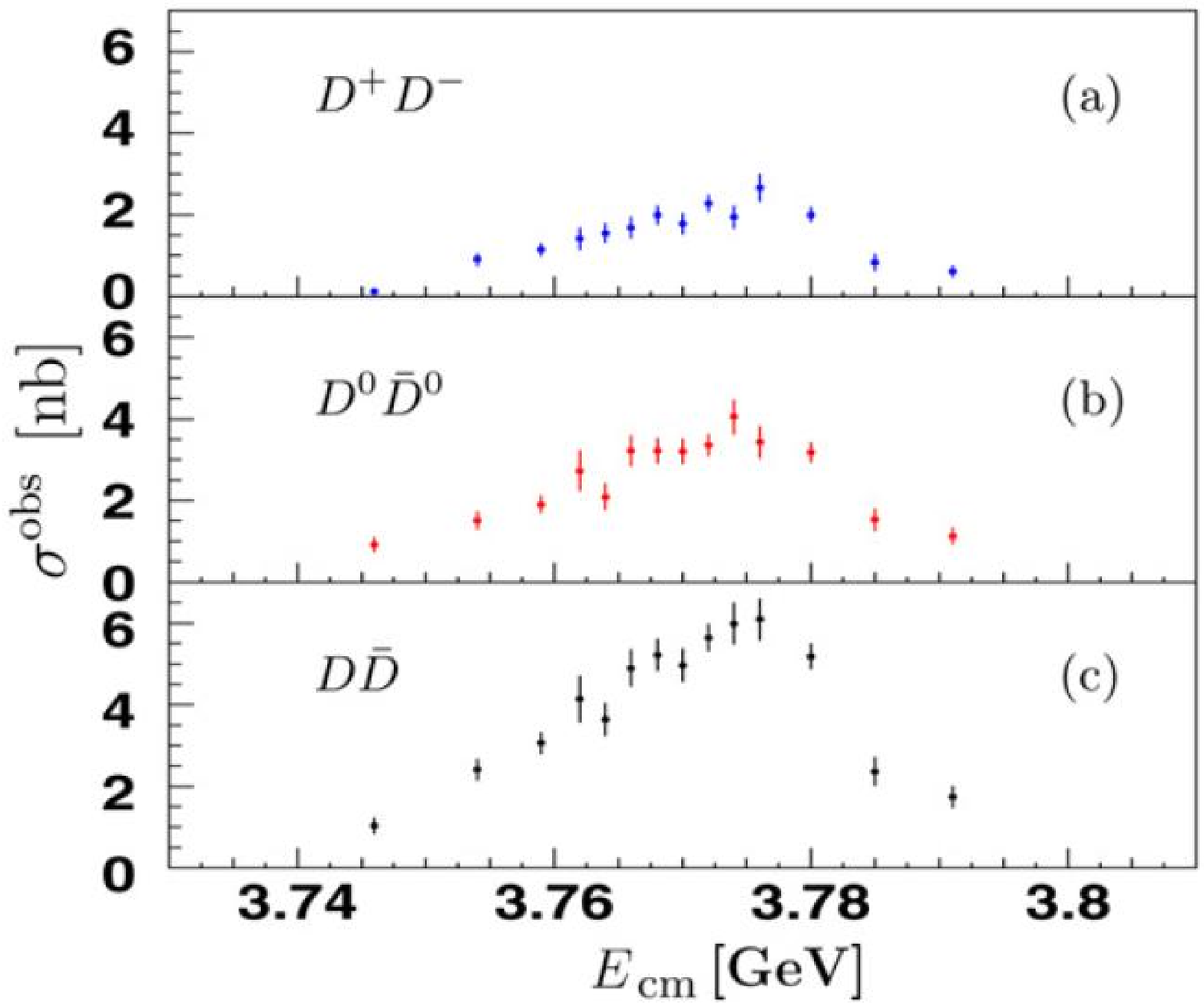}
\caption{The observed cross sections for
(a) $e^+e^-\to D^0\bar D^0$,
(b) $e^+e^-\to D^+D^-$ and
(c) $e^+e^-\to D\bar D$ versus the nominal center-of-mass energies.}
\label{fig:crs_dd}
\end{figure*}

\begin{figure*}[hbtp]
\centering
\includegraphics[width=8cm,height=6cm]{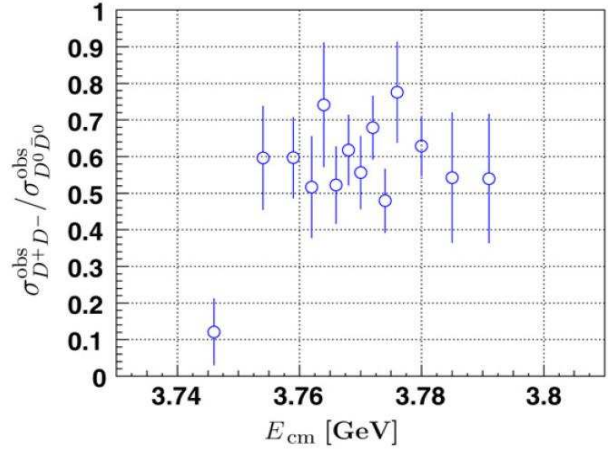}
\caption{The ratio of the observed cross section for
$e^+e^-\to D^+D^-$ relative to the observed cross section for
$e^+e^-\to D^0\bar D^0$ versus the nominal center-of-mass
energies.}
\label{fig:ratio_dd}
\end{figure*}

These measurements are made by analyzing the precision cross section
data sets taken in March 2001, during the period from March to April
2003, and during December 2003 to January 2004.
Fig. \ref{fig:crs_dd} shows the observed cross sections for $e^+e^-\to
D^0\bar D^0$, $e^+e^-\to D^+D^-$ and $e^+e^-\to D\bar D$ versus the
nominal center-of-mass energies. In the figure, we can see that the line
shapes of the cross section for $e^+e^-\to D^0\bar D^0$, $e^+e^-\to D^+D^-$
and $e^+e^-\to D\bar D$ are also anomalous, just like the line shape of
the observed cross section for $e^+e^-\to$ hadrons.
Fig. \ref{fig:ratio_dd} shows the measured ratio of the observed cross
section for $e^+e^-\to D^+D^-$ relative to the observed cross section for
$e^+e^-\to D^0\bar D^0$ versus the nominal center-of-mass energies.

\section{Search for $\psi(3770)\to$ exclusive light hadron
\cite{to_be_submit1,to_be_submit2}}
We measure the observed cross sections for the exclusive light hadron
final states of $\pi^+\pi^-\pi^0\pi^0$, $K^+K^-\pi^0\pi^0$,
$2(\pi^+\pi^-\pi^0)$, $K^+K^-\pi^+\pi^-\pi^0\pi^0$,
$3(\pi^+\pi^-)\pi^0\pi^0$, $K_S^0K^-\pi^+$,
$K_S^0K^-\pi^+\pi^0$, $K_S^0K^-\pi^+\pi^+\pi^-$,
$K_S^0K^-\pi^+\pi^+\pi^-\pi^0$,
$K_S^0K^-\pi^+\pi^+\pi^+\pi^-\pi^-$ and
$K_S^0K^-\pi^+\pi^0\pi^0$
at the center-of-mass energies of $\sqrt s=3.773$, 3.650 and 3.6648 GeV.
The preliminary results are shown in Tab. \ref{tab:crs_light_hadrons},
where the upper limits are set at 90\% C.L..
By comparing the observed cross sections for each process
measured at $\sqrt s= 3.773$ and 3.650 GeV, we set the upper limits,
$B^{\rm up}_{\psi(3770)\to f}$, on the branching fractions for
$\psi(3770)$ decay to these final states at $90\%$ C.L..
These measurements provide helpful
information to understand the mechanism of the continuum light hadron
production and the discrepancy between the observed cross sections for
$D\bar D$ and $\psi(3770)$ production.

In the measurements, we ignore the interference effects between the
continuum and resonance amplitudes due to not knowing the details
about the two amplitudes, and neglect the difference of the
vacuum polarization corrections at the different energy points.
In this case, up to now, we still can not draw a conclusion that the
$\psi(3770)$ does not decay into these final states even if we have not
observed significant difference between the observed cross sections
for $e^+e^-\to$ most light hadron final states at the two energy points.
We are looking forward to extract the branching fractions for $\psi(3770)\to$
exclusive light hadron by analyzing large data samples taken at more different
energy points covering both $\psi(3770)$ and $\psi(3686)$ \cite{crsscan}.

\begin{table*}[htbp]
\begin{center}
\caption{
The measured observed cross sections for $e^+e^-\to$ exclusive
light hadrons at $\sqrt s=$ 3.773, 3.650 and 3.6648 GeV, and
the upper limits on the branching fractions for $\psi(3770)$ decay into
these final states.
\label{tab:crs_light_hadrons}}
\begin{tabular}{|l|c|c|c|c|} \hline
Final State   & $\sigma^{\rm up}_{e^+e^-\to f}$(@3.773 GeV)
              & $\sigma^{\rm up}_{e^+e^-\to f}$(@3.650 GeV)
              & $\sigma^{\rm up}_{e^+e^-\to f}$(@3.6648 GeV)
              & $B^{\rm up}_{\psi(3770)\to f}$ \\
&[pb]&[pb]&[pb]&$\times$10$ ^{-3}$\\ \hline
$\pi^+\pi^-\pi^0\pi^0$             &$214.8\pm25.0\pm27.9$   &$292.2\pm43.4\pm39.4$   
&$397.3\pm91.3\pm50.9$&$<8.9$\\
$K^+K^-\pi^0\pi^0$                 &$26.7\pm12.4\pm4.0$     &$54.3\pm22.1\pm8.0$     
&$<161.7$&$<4.2 $\\
$2(\pi^+\pi^-\pi^0)$               &$1051.5\pm119.2\pm148.3$&$1077.3\pm178.1\pm164.8$
&$876.4\pm296.1\pm121.8$&$<58.5$\\
$K^+K^-\pi^+\pi^-\pi^0\pi^0$       &$186.0\pm86.4\pm31.2$   &$379.1\pm139.1\pm57.2$  
&$<926.2$&$<26.7$\\
$3(\pi^+\pi^-)\pi^0\pi^0$          &$973.8\pm262.8\pm180.2$ &$842.1\pm311.1\pm160.8$ 
&$<2623.1$&$<117.4$\\
\hline
$K_S^0K^-\pi^+$                    &$15.2\pm3.8\pm1.6$&$1.5\pm2.3\pm0.2$&-&$<3.1$\\
$K_S^0K^-\pi^+\pi^0$               &$96.2\pm15.9\pm11.1$&$47.9\pm18.0\pm5.6 $&-&$<13.3$\\
$K_S^0K^-\pi^+\pi^+\pi^-$          &$91.5\pm15.3\pm13.0$&$81.4\pm23.1\pm11.5 $&-&$<8.7$\\
$K_S^0K^-\pi^+\pi^+\pi^-\pi^0$     &$253.0\pm57.1\pm38.4$&$119.0\pm64.7\pm20.5$&-&$<41.8$\\
$K_S^0K^-\pi^+\pi^+\pi^+\pi^-\pi^-$&$44.4\pm21.9\pm8.2$&$<69.7$&-&$<12.2$\\
$K_S^0K^-\pi^+\pi^0\pi^0$          &$147.0\pm42.4\pm21.0$&$67.1\pm40.7\pm9.5$&-&$<26.5$\\
\hline
\end{tabular}
\end{center}
\end{table*}

\section{Summary}

Using the $e^+e^-$ data sets of about 33, 6.5 and 1.0 pb$^{-1}$,
respectively, taken around the center-of-mass energies of $\sqrt s=$
3.773 GeV, at $\sqrt s=$ 3.650 GeV and at $\sqrt s=$ 3.6648 GeV with the BES-II
detector at the BEPC collider, BES Collaboration measure the
branching fraction for $\psi(3770)\to$ non-$D\bar D$ decays by analyzing several
different data samples and different methods with assumption that there is
only one simple $\psi(3770)$ in the energy region between 3.700 and 3.872
GeV. We observe an anomalous line shape of the cross section for $e^+e^-\to$
hadrons in the energy region from 3.650 to 3.872 GeV. We measure the line
shapes of the $D^+D^-$, $D^0\bar D^0$ and $D\bar D$ production and the
ratio of the production rates of $D^+D^-$ and $D^0\bar D^0$ in $e^+e^-$
annihilation at $\psi(3770)$ resonance. These indicate that there may exist
a new structure in addition to one simple $\psi(3770)$ resonance in the
energy region between 3.700 and 3.872 GeV or there are some unknown dynamics
effects distorting the line shape of the cross sections for
$e^+e^-\to$ hadrons and $D\bar D$. We also search for $\psi(3770)$
decay into exclusive light hadron processes containing $K^0_S$ or
$\pi^0\pi^0$ mesons in the final states.

\begin{acknowledgments}
The BES collaboration thanks the staff of BEPC for their hard
efforts. This work is supported in part by the National Natural
Science Foundation of China under contracts Nos. 10491300,
10225524, 10225525, 10425523, the Chinese Academy of Sciences
under contract No. KJ 95T-03, the 100 Talents Program of CAS under
Contract Nos. U-11, U-24, U-25, the Knowledge Innovation Project
of CAS under Contract Nos. U-602, U-34 (IHEP), the National
Natural Science Foundation of China under Contract  No. 10225522
(Tsinghua University).
\end{acknowledgments}


\begin{thebibliography}{9}   

\bibitem{rzhc}
G. Rong, D. H. Zhang, J. C. Chen, hep-ex/0506051.

\bibitem{prl96_092002}
CLEO Collaboration, D. Besson, et al., Phys. Rev Lett. {\bf 96} (2006) 092002.

\bibitem{plb641_145} BES Collaboration, M. Ablikim et al.,
Phys. Lett. {\bf B 641} (2006) 145.

\bibitem{prl97_121801} BES Collaboration, M. Ablikim et al.,
Phys. Rev. Lett. {\bf 97} (2006) 121801.

\bibitem{prd76_122002} BES Collaboration, M. Ablikim et al.,
Phys. Rev. {\bf D 76} (2007) 122002.

\bibitem{plb659_74} BES Collaboration, M. Ablikim et al.,
Phys. Lett. {\bf B 659} (2008) 74.

\bibitem{pdg2008}  Particle Data Group, C. Amsler, et al.,
Phys. Lett. {\bf B 667} (2008) 1.

\bibitem{hepnp28_325} BES Collaboration, J. Z. Bai et al.,
HEP $\&$ NP {\bf 28(4)} (2004) 325.

\bibitem{plb605_63} BES Collaboration, M.Ablikim et al.,
Phys. Lett. {\bf B 605} (2005) 63.

\bibitem{prl96_082004} CLEO Collaboration, N. E. Adam et al., Phys.
Rev. Lett. {\bf 96} (2006) 082004.

\bibitem{prl96_182002} CLEO Collaboration, T. E. Coans et al., Phys.
Rev. Lett. {\bf 96} (2006) 182002.

\bibitem{prd74_031106} CLEO Collaboration, B. A. Briere et al.,
Phys. Rev. {\bf D 74} (2006) 031106.

\bibitem{prd74_012005} CLEO Collaboration, D. Cronin-Hennessy et al.,
Phys. Rev. {\bf D 74} (2006) 012005.

\bibitem{plb650_111} BES Collaboration, M. Ablikim et al.,
Phys. Lett. {\bf B 650} (2007) 111.

\bibitem{plb656_30} BES Collaboration, M. Ablikim et al.,
Phys. Lett. {\bf B 656} (2007) 30.

\bibitem{epjc52_805} BES Collaboration, M. Ablikim et al.,
Eur. Phys. J {\bf C 52} (2007) 805.

\bibitem{prl101_102004} BES Collaboration, M. Ablikim et al.,
Phys. Rev. Lett. {\bf 101} (2008) 102004.

\bibitem{babar} BABAR Collaboration, B. Aubert et al.,
arXiv:hep-ex/0710.1371v1.

\bibitem{belle} BELLE Collaboration, G. Pakhlovs et al.
arXiv:hep-ex/0708.0082v2.

\bibitem{plb668_263} BES Collaboration, M. Ablikim et al.,
Phys. Lett. {\bf B 668} (2008) 263.

\bibitem{to_be_submit1} BES Collaboration, M. Ablikim et al.,
to be submitted.

\bibitem{to_be_submit2} BES Collaboration, M. Ablikim et al.,
to be submitted.

\bibitem{crsscan} G. Rong (For BES Collaboration),
Int. J. Mod. Phys. {\bf A 21} (2006) 5416.

\end{thebibliography}
\end{document}